\documentclass[12pt]{article}
\usepackage{amsmath}
\usepackage{epsfig}
\usepackage{amssymb}
\input{epsf}
\newcommand\as{\alpha_s}
\newcommand\f[2]{\frac{#1}{#2}}

\def\beq{\begin{equation}}
\def\eeq{\end{equation}}
\def\beeq{\begin{eqnarray}}
\def\eeeq{\end{eqnarray}}
\def\to{\rightarrow}
\def\nn{\nonumber}

\def\b0{b_0}

\begin{document}

\begin{titlepage}
\renewcommand{\thefootnote}{\fnsymbol{footnote}}
\begin{flushright}
BNL-NT-05/20 \\
RBRC-527 \\
hep-ph/0506150
     \end{flushright}
\par \vspace{10mm}
\begin{center}
{\Large \bf
Threshold resummation for the \\[5mm] prompt-photon 
cross section revisited}

\end{center}
\par \vspace{2mm}
\begin{center}
{\bf Daniel de Florian${}^{\,a}$}
\hskip .2cm
and
\hskip .2cm
{\bf Werner Vogelsang$
{}^{\,b}$}\\

\vspace{5mm}
${}^{a}\,$Departamento de F\'\i sica, FCEYN, Universidad de Buenos Aires,\\
(1428) Pabell\'on 1 Ciudad Universitaria, Capital Federal, Argentina

${}^{b}\,$Physics Department and RIKEN-BNL Research Center, \\
Brookhaven National Laboratory, Upton, NY 11973, U.S.A.\\

\end{center}

\par \vspace{9mm}
\begin{center} {\large \bf Abstract} \end{center}
\begin{quote}
\pretolerance 10000
We study the resummation of large logarithmic perturbative corrections
to the partonic cross sections relevant for the process
$pp\to \gamma X$ at high transverse momentum of the photon.
These corrections arise near the threshold for the partonic
reaction and are associated with soft-gluon emission. We especially 
focus on the resummation effects for the contribution to the
cross section where the photon is produced in jet fragmentation.
Previous calculations in perturbation theory at fixed-order 
have established that this contribution is a subdominant part of the
cross section. We find, however, that it is subject to much larger 
resummation effects than the direct (non-fragmentation) piece and 
therefore appears to be a significant contribution in the
fixed-target regime, not much suppressed with respect to the 
direct part. Inclusion of threshold resummation for the 
fragmentation piece leads to some improvement in comparisons 
between theoretical calculations and experimental data.

\end{quote}

\end{titlepage}

\setcounter{footnote}{1}
\renewcommand{\thefootnote}{\fnsymbol{footnote}}

\noindent
{\bf Introduction.} 
Prompt-photon production at high transverse momentum~\cite{photondata}, 
$pp,p\bar{p},pN\rightarrow \gamma X$, has been a classic tool for 
constraining the nucleon's gluon density, because at leading order a photon 
can be produced in the Compton reaction $qg\to\gamma q$. The ``point-like'' 
coupling of the photon to the quark provides a potentially clean 
electromagnetic probe of QCD hard scattering. However, a pattern of 
disagreement between theoretical next-to-leading order (NLO)
predictions~\cite{nlogam,nlofrag} and experimental data~\cite{cdf,e706,ua6}
for prompt photon production has been observed in recent 
years, not globally curable by ``fine-tuning'' the gluon 
density~\cite{vv,huston95,aurenche99}. The most serious problems 
relate to the fixed-target regime, where NLO theory shows a dramatic shortfall
when compared to some of the data sets~\cite{e706,ua6}. We note that the 
mutual consistency of the data sets has been questioned~\cite{aurenche99}. 
Nevertheless, for the related single-inclusive neutral-pion production, 
$pp\rightarrow \pi^0 X$, comparisons between NLO calculations and 
data from mostly the same experiments have also shown a systematic 
disagreement~\cite{aur,baur}.

In a recent paper~\cite{ddfwv}, we have shown that a drastic 
improvement of the theoretical description of single-inclusive pion
production in the fixed-target regime is found when certain 
large perturbative contributions to the partonic hard-scattering
cross sections are taken into account to all orders in perturbation
theory. These terms, known as threshold logarithms,  
arise near partonic threshold, when the initial partons have
just enough energy to produce a high-transverse momentum
parton (which subsequently fragments into the observed hadron)
and a massless recoiling jet. In this case, the phase space available 
for gluon radiation vanishes, resulting in large logarithmic corrections to
the partonic cross section. For the cross section integrated over
all rapidities, the most important (``leading'') 
logarithms are of the form $\as^k \ln^{2k}\left(1-\hat{x}_T^2\right)$ 
at the $k$th order in perturbation theory, where $\as$ is the 
strong coupling and $\hat{x}_T\equiv 2 p_T/\sqrt{\hat{s}}$, with $p_T$ the 
parton transverse momentum and $\sqrt{\hat{s}}$ the partonic 
center-of-mass (cms) energy. Sufficiently close to threshold, NLO, which 
captures only the term for $k=1$, will not be adequate anymore; instead,
all logarithmic terms will become relevant and thus need to be
taken into account. This is achieved by threshold 
resummation~\cite{dyresum,KS,BCMN}.

The improvement of the comparison between data and theory
due to threshold resummation in pion production has motivated
us to revisit prompt-photon production. Here, too, 
large logarithmic corrections arise near partonic threshold. 
There is an extensive earlier literature~\cite{LOS,CMN,CMNOV,KO,sv,lsv} 
on QCD resummations for the ``direct'' partonic processes 
$qg\to \gamma q$ and $q\bar{q}\to \gamma g$.
The corresponding phenomenological studies for threshold
resummation~\cite{CMNOV,KO,sv} have 
found only a relatively small enhancement of the theoretical
prediction by threshold resummation, not generally sufficient to 
provide satisfactory agreement with the fixed-target prompt-photon
data. In the present paper, we will extend the previous 
studies of threshold resummation effects in prompt-photon production 
by including also the resummation for the ``fragmentation'' 
component in the cross section, to which we turn now. 

As is well-known~\cite{owe87} (for discussion 
and references, see also Ref.~\cite{photondata}), 
high-$p_T$ photons are not only produced
by the ``direct'' contributions from the partonic hard processes
$ab\to \gamma c$, but also in
jet fragmentation, when a parton $f$ emerging from the hard-scattering
process fragments into a photon plus a number of hadrons.
The need for introducing a fragmentation contribution is physically
motivated from the fact that a fragmentation process may produce,
for example, a $\rho$ meson that converts into a photon,
leading to the same signal. In addition, at higher orders, the
perturbative direct component contains divergencies from configurations 
where a final-state quark becomes collinear to the photon. These 
long-distance contributions naturally introduce the need for 
non-perturbative fragmentation functions $D^{f\to \gamma}$
into which they can be absorbed. The fragmentation component
is of the same perturbative order as the direct one, 
${\cal O}(\alpha_{em}\alpha_s)$, since the underlying lowest-order (LO)
partonic processes are the ${\cal O}(\alpha_s^2)$ QCD
scatterings $ab\to fc$, and the fragmentation functions $D^{f\to \gamma}$ 
are of order $\alpha_{em}/\alpha_s$ in QCD.
There is some knowledge about the photon fragmentation
functions from the LEP experiments~\cite{buskulic96a}. 
Theoretical model predictions~\cite{aurf,grvf,bourhis98} 
for the photon fragmentation functions are compatible with these data.
Using these sets of $D^{f\to \gamma}$, one can then estimate 
the fragmentation contribution to the prompt photon cross
section. NLO calculations in the fixed-target regime show~\cite{photondata,vv}
that fragmentation photons contribute about $10-30\%$ to the 
prompt-photon photon cross section. Here, the precise value depends both
on the photon transverse momentum, but also on the type of hadron beams 
used. Generally, because of the additional fragmentation function
and because of the different underlying hard-scattering
processes, the fragmentation component is suppressed and also expected 
to fall off more rapidly in $p_T$ than the direct one. On the other 
hand, in $pp$ or $pN$ (as opposed to $p\bar{p}$) collisions, the direct 
channels $qg\to \gamma q$ and $q\bar{q}\to \gamma g$ always involve 
either a sea quark or gluon distribution in the initial state, which 
both decrease rapidly towards larger momentum fractions, leading
to a rapid decrease of the cross section at high $p_T$. In contrast, 
the fragmentation piece has contributions from $qq$ scattering~\cite{CMNOV},
involving two valence densities. As a result, for $pp$ or $pN$ 
collisions, the fragmentation component may continue to be
sizable relative to the direct part out to quite high transverse
momenta.

Despite the fact that according to the NLO calculation the 
fragmentation contribution is only a subdominant part of the 
cross section, in the light of the results of Ref.~\cite{ddfwv}
it deserves a closer investigation. There, as we mentioned 
above, very large enhancements were found for $pp\to \pi^0 X$ 
in the fixed-target regime. In the theoretical calculation,
the only difference between $pp\to \pi^0 X$ and the fragmentation
component to $pp\rightarrow \gamma X$ is the use of different
fragmentation functions. One therefore expects that also
for the fragmentation component to prompt photon production
there could be a large increase from resummation. Since
it is known from the previous studies~\cite{CMNOV,KO,sv}
that the direct part receives only moderate resummation
effects, it is likely that the relative importance
of the fragmentation contribution in the fixed-target regime
is actually much larger than previously estimated on the
basis of the NLO calculations. The precise details
will of course depend on the photon fragmentation functions.
The $D^{f\to \gamma}$ are much more peaked at large momentum
fractions $z$ than pion fragmentation functions, due to the
perturbative (``point-like'') piece in the 
evolution~\cite{aurf,grvf,bourhis98}. 
On the other hand, the gluon fragmentation function will 
be relatively much less important than in the pion case,
meaning that some partonic channels with large resummation
effects, such as $gg\to gg$, are less important. In the present
paper, we present a brief phenomenological study of the
resummation effects for the fragmentation part of the
prompt photon cross section, and their implications for the
comparison with the fixed-target data. Irrespective of how
well theory and fixed-target data sets agree after the 
resummation of the fragmentation part is included, 
the latter is an important ingredient of the theoretical
calculation of the cross section. 

\noindent
{\bf Resummed cross section.} The cross section for 
$H_1 H_2 \rightarrow \gamma X \,$ may be written as
\begin{align}
\label{eq:1} \f{p_T^3\, d\sigma(x_T)}{dp_T} = \sum_{a,b,f}\, &
\int_0^1 dx_1 \, f_{a/H_1}\left(x_1,\mu^2\right) \, \int_0^1
dx_2 \, f_{b/H_2}\left(x_2,\mu^2\right) \, \int_0^1 dz
\,z^2\, D^{f\to \gamma}\left(z,\mu^2\right) \, \nn \\ &\int_0^1
d\hat{x}_T \, \, \delta\left(\hat{x}_T-\f{x_T}{z\sqrt{x_1
x_2}}\right) \, \int_{\hat{\eta}_{-}}^{\hat{\eta}_{+}} d\hat{\eta}
\, \f{\hat{x}_T^4 \,\hat{s}}{2} \,
 \f{d\hat{\sigma}_{ab\rightarrow fX}(\hat{x}_T^2,\hat{\eta},\mu)}
{ d\hat{x}_T^2 d\hat{\eta}} \, .
\end{align}
We have integrated over all pseudorapidities $\eta$ of the produced
photon. $\hat{\eta}$ is the pseudorapidity at parton level, with
$\hat{\eta}_{+}=-\hat{\eta}_{-}=\ln\left[(1+\sqrt{1-\hat{x}_T^2})/
\hat{x}_T\right]$. The sum in Eq.~(\ref{eq:1}) runs over all
partonic subprocesses $ab\to fX$, with partonic cross sections
$d\hat{\sigma}_{ab\rightarrow fX}$, parton distribution functions
$f_{a/H_1}$ and $f_{b/H_2}$, and parton-to-photon fragmentation
functions $D^{f\to \gamma}$. The direct contributions are included 
and are obtained by setting $f=\gamma$ and $D^{f\to \gamma}=\delta(1-z)$. 
$\mu$ denotes the factorization/renormalization scales, which we
have chosen to be equal for simplicity. 

The partonic cross sections are computed in QCD perturbation theory.
Their expansions begin at ${\cal O}(\as \alpha_{em})$ for the direct
part, and at ${\cal O}(\as^2)$ for the fragmentation part. Defining
\begin{equation}
\Sigma_{ab\rightarrow fX}(\hat{x}_T^2,\mu)\equiv
\int_{\hat{\eta}_{-}}^{\hat{\eta}_{+}} d\hat{\eta} \,
\f{\hat{x}_T^4 \,\hat{s}}{2} \, \f{d\hat{\sigma}_{ab\rightarrow
fX}(\hat{x}_T^2,\hat{\eta},\mu)}{ d\hat{x}_T^2 d\hat{\eta}} \, ,
\end{equation}
one finds at NLO the structure~\cite{nlogam,nlofrag}
\begin{eqnarray}
\label{eq:sigma1}
\Sigma_{ab\rightarrow fX}(\hat{x}_T^2,\mu)=
\Sigma_{ab\rightarrow fX}^{{\rm (Born)}}(\hat{x}_T^2)\,
\left[ 1+\alpha_s(\mu)\left\{ A \ln^2(1-\hat{x}_T^2)+
B \ln(1-\hat{x}_T^2)+C+\ldots\right\} \right] \; ,
\end{eqnarray}
where $\Sigma_{ab\rightarrow fX}^{{\rm (Born)}}$ is the 
Born cross section for the process $ab\to fX$, and $A,B,C$
are coefficients that depend on the partonic process. Finally,
the ellipses denote terms that vanish at $\hat{x}_T=1$.
The logarithmic terms are the leading and next-to-leading 
logarithms (LL, NLL) at this order. At higher orders, the 
logarithmic contributions are enhanced by terms proportional to 
$\as^k \, \ln^m (1-\hat{x}_T^2)$, with $m\leq 2k$, at the $k$th 
order of $\Sigma_{ab\rightarrow fX}$. As we discussed earlier, these 
logarithmic terms are due to soft-gluon radiation and may be
resummed to all orders in $\alpha_s$. The resummation
discussed in this work deals with the ``towers'' for $m=2k,2k-1,2k-2$.

As follows from Eq.~(\ref{eq:1}), since the observed
$x_T=2p_T/\sqrt{S}$ is fixed, $\hat{x}_T$ assumes particularly
large values when the partonic momentum fractions approach
the lower ends of their ranges. Since the parton distributions
rise steeply towards small argument,
this generally increases the relevance of the threshold
regime, and the soft-gluon effects are relevant even for situations
where the the hadronic center-of-mass energy is much larger than the
transverse momentum of the final state hadrons. This effect, valid in
general in hadronic collisions, is even enhanced in the fragmentation 
contribution since only a fraction $z$ of the available energy
is actually used to produce the final-state photon.

The resummation of the soft-gluon contributions is carried out in
Mellin-$N$ moment space, where the convolutions in Eq.~(\ref{eq:1})
between parton
distributions, fragmentation functions, and subprocess cross sections
factorize into ordinary products. We take Mellin moments in the scaling
variable $x_T^2$ as
\begin{align}
\label{eq:moments}
\sigma(N)\equiv \int_0^1 dx_T^2 \, \left(x_T^2 \right)^{N-1} \;
\f{p_T^3\, d\sigma(x_T)}{dp_T} \, .
\end{align}
In $N$-space Eq.(\ref{eq:1}) becomes
\begin{align}
\sigma(N)=\sum_{a,b,f} \,  f_{a/H_1}(N+1,\mu^2) \,
f_{b/H_2}(N+1,\mu^2) \,  D^{f\to \gamma}(2N+3,\mu^2) \,
\Sigma_{ab\rightarrow fX}(N)\, ,
\end{align}
with the usual Mellin moments of the parton distribution functions
and fragmentation functions. As before, for the direct
contributions, one has $D^{f\to \gamma}=\delta(1-z)$ and
therefore $D^{f\to \gamma}(2N+3,\mu^2)=1$. In addition,
\begin{align} \label{momdef}
\Sigma_{ab\rightarrow fX}(N) \equiv \int_0^1 d\hat{x}_T^2 \,
\left(\hat{x}_T^2 \right)^{N-1}\, \Sigma_{ab\rightarrow
fX}(\hat{x}_T^2)\,.
\end{align}
Here, the threshold limit $\hat{x}_T^2\to 1$ corresponds 
to $N\to \infty$, and the leading soft-gluon corrections arise as 
terms $\propto \as^k \ln^{2k}N$.

In Mellin-moment space, threshold resummation results in
exponentiation of the soft-gluon corrections. In case of 
a single-inclusive cross section, the structure of the resummed 
result reads for a given partonic channel~\cite{ddfwv,CMN,KOS,KO1}
\begin{align}
\label{eq:res}
\Sigma^{{\rm (res)}}_{ab\to cd} (N-1)=  C_{ab\to cd}\,
\Delta^a_N\, \Delta^b_N\, \Delta^c_N\,
J^{d}_N\, \left[ \sum_{I} G^{I}_{ab\to cd}\,
\Delta^{{\rm (int)} ab\rightarrow cd}_{I\, N}\right] \,
\Sigma^{{\rm (Born)}}_{ab\to cd} (N-1) \;  .
\end{align}
Each of the functions $\Delta_N^{a,b,c}$, $J_N^d$, $\Delta^{{\rm (int)} 
ab\rightarrow cd}_{I\, N}$ is an exponential. The
$\Delta^{a,b,c}_N$ represent the effects of
soft-gluon radiation collinear to initial partons $a,b$ or the
``observed'' final-state parton $c$. The function $J^{d}_N$
embodies collinear, soft or hard, emission by the non-observed
parton $d$. Large-angle soft-gluon emission is accounted for by the factors
$\Delta^{{\rm (int)} ab\rightarrow cd}_{I\, N}$, which depend on
the color configuration $I$ of the participating partons.
The sum runs over all possible color configurations $I$,
with $G^{I}_{ab\to cd}$ representing a weight for
each color configuration, such that $\sum_I G^{I}_{ab\to cd}=1$.
Finally, the coefficient $C_{ab\to cd}$ contains
$N$-independent hard contributions arising from one-loop
virtual corrections.

The explicit NLL expressions for all the factors in Eq.~(\ref{eq:res})
may be found in Refs.~\cite{ddfwv,CMN}. The factors 
$\Delta_N^{a,b,c}$ and $J_N^d$ contain the leading 
logarithms and are universal 
in the sense that they only depend on the color 
charge of the parton they represent. Eq.~(\ref{eq:res})
applies to the direct as well as to the fragmentation component. 
In the former, the ``observed'' parton is the photon, and 
thus $\Delta_N^c=1$. Also, in this case 
there is only one color structure of the hard scattering,
so that the sum in Eq.~(\ref{eq:res}) contains only
one term. In contrast, several color channels contribute to
each of the $2\to 2$ QCD subprocesses relevant for the fragmentation
part. As a result, there are color interferences and correlations
in large-angle soft-gluon emission at NLL, and the resummed cross
section for each subprocess becomes a sum of exponentials, rather
than a single one. The complete expressions for the 
$\Delta^{{\rm (int)} ab\rightarrow cd}_{I\, N}$, $G^{I}_{ab\to cd}$ 
and $C_{ab\to cd}$ are also given in Ref.~\cite{CMN} for the 
direct case, and in~\cite{ddfwv} for the fragmentation 
part. 

In the resummed exponent, the large logarithms in $N$
occur only as single logarithms, of the form
$\as^k \ln^{k+1}(N)$ for the leading terms. 
Subleading terms are down by one or more powers of $\ln(N)$. 
Soft-gluon effects are partly already contained in 
the ($\overline{{\mathrm{MS}}}$-defined) parton distribution functions 
and fragmentation functions. As a result, it turns out that they enhance 
the cross section~\cite{dyresum,CMN}. We also note that the 
factors $\Delta_N^i$ depend on the factorization scale in such a way
that they will compensate the scale dependence (evolution)
of the parton distribution and fragmentation functions. One
therefore expects a decrease in scale dependence of the 
predicted cross section~\cite{BCMN,Catani:1996yz,sv1}.

We finally note that from the large Mellin-$N$ point of view
the fragmentation component is at first sight suppressed by 
$1/N$~\cite{CMNOV} since the photon fragmentation functions always involve a 
``quark-to-photon'' splitting function $P_{\gamma q}$ 
which in moment space is $\propto 1/N$. However, as was
pointed out in~\cite{CMNOV}, this suppression may be compensated
in particular for $pp$ or $pN$ collisions by the fact that the 
fragmentation component involves quark-quark scattering, whereas
the direct piece proceeds through quark-antiquark or quark-gluon
scattering (see above). At large $N$, the quark channels with their
valence component dominate. In any case, the resummed corrections 
for the fragmentation component constitute by themselves a 
well-defined set of higher-order corrections which has much phenomenological
relevance as we will see below. That said, we emphasize that 
a more detailed analysis of $1/N$-suppressed contributions
also in the direct part would be desirable for future work.

\noindent
{\bf Phenomenological results.} We will now present some
phenomenological results for the prompt photon cross 
section, taking into account the resummation for both
the direct and the fragmentation parts. 
This is not meant to be an exhaustive study of
the available data for direct-photon production; rather
we should like to investigate the overall size and relevance
of the resummation effects and in particular the question
in how far they change the relative importance of direct
and fragmentation contributions. We therefore select
only a few representative data sets to compare to:
the E706 data for prompt-photon production in $p\,\!Be$ 
scattering~\cite{e706} at $\sqrt{S}=31.5$~GeV, the $pp$ data 
from UA6~\cite{ua6} ($\sqrt{S}=24.3$~GeV), and the  
data from R806~\cite{r806} taken in $pp$ collisions at
the ISR at $\sqrt{S}=63$~GeV.

In order to obtain a resummed cross section in $x_T^2$ space, 
one needs an inverse Mellin transform. 
As previous studies~\cite{ddfwv,CMNOV,sv} we will use the 
``Minimal Prescription'' developed in Ref.~\cite{Catani:1996yz},
for which one chooses a Mellin contour in complex-$N$ space that lies 
to the {\it left} of the poles at $\lambda=1/2$ and $\lambda=1$ in the 
resummed Mellin integrand, where $\lambda=\alpha_s(\mu^2) b_0
\ln (N)$ with $b_0=( 33 - 2 N_f)/12\pi$, but to the 
right of all other poles. 

When performing the resummation, one of course wants to make full
use of the available fixed-order cross section, which in our case
is NLO (${\cal O}(\alpha_{em}\as^2)$). Therefore, a matching to this cross
section is appropriate, which may be achieved by expanding the resummed
cross section to NLO, subtracting the expanded result
from the resummed one, and adding the ``exact'' NLO cross 
section~\cite{nlogam,nlofrag}:
\begin{align}
\label{hadnres}
\f{p_T^3\, d\sigma^{\rm (match)}(x_T)}{dp_T} &= \sum_{a,b,f}\,
\;\int_{\cal C}
\;\frac{dN}{2\pi i} \;\left( x_T^2 \right)^{-N}
\; f_{a/h_1}(N+1,\mu^2) \; f_{b/h_2}(N+1,\mu^2) \;
D^{f\to \gamma}(2N+3,\mu^2)
 \nn \\
&\times \left[ \;
\Sigma^{\rm (res)}_{ab\to fd} (N)
- \left. \Sigma^{{\rm (res)}}_{ab\to fd} (N)
\right|_{{\rm NLO}} \, \right]
+\f{p_T^3\, d\sigma^{\rm (NLO)}(x_T)}{dp_T}
 \;\;,
\end{align}
where $\Sigma^{{\rm (res)}}_{ab\to cd} (N)$ is the resummed cross
section for the partonic channel $ab\to cd$ as given in Eq.~(\ref{eq:res}).
In this way, NLO is taken into account in full, and the soft-gluon
contributions beyond NLO are resummed to NLL. Any double-counting
of perturbative orders is avoided. Note that, as before, this cross 
section is the sum of both direct and fragmentation contributions.

As we have discussed earlier, we perform the resummation
for the fully rapidity-integrated cross section. In experiment always
only a certain limited range of rapidity is covered. In order to be able
to compare to data, we therefore approximate the cross section in
the experimentally accessible rapidity region by~\cite{ddfwv,CMNOV}
\begin{equation}
\f{p_T^3\, d\sigma^{\rm (match)}}{dp_T}({\rm \eta\, in\, exp.\, range})
= \f{d\sigma^{\rm (match)}({\rm all \, \eta})}
{d\sigma^{\rm (NLO)}({\rm all \, \eta})} \; 
\f{p_T^3\, d\sigma^{\rm (NLO)}}{dp_T}
({\rm \eta\, in\, exp.\, range})\, .
\end{equation}
In other words, we rescale the matched resummed result by the ratio of
NLO cross sections integrated over the experimentally relevant
rapidity region or over all $\eta$, respectively.

Our choice for the parton distribution functions
will be the CTEQ6 set~\cite{cteq6}. For the photon fragmentation 
functions we use those of~\cite{grvf}. We note that other
sets have been proposed~\cite{aurf,bourhis98} for the latter. 

\begin{figure}[t!]
\begin{center}
\epsfig{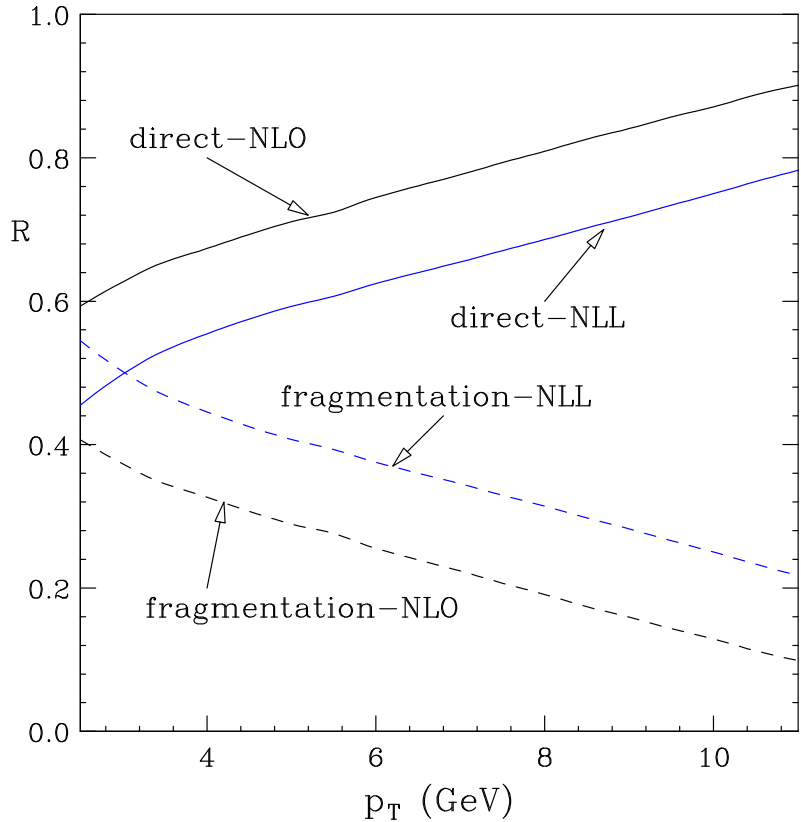}
\end{center}
\vspace*{-.5cm}
\caption{Relative contributions of direct and fragmentation
photons to the cross section at NLO and for the NLL resummed
case, for $pp$ collisions at $\sqrt{S}=31.5$~GeV. We have chosen
all scales as $\mu=p_T$. 
\label{fig1} }
\vspace*{0.cm}
\end{figure}

We start by comparing the relative importance of 
the photon fragmentation contribution at NLO and after
NLL resummation of the threshold logarithms. Figure~\ref{fig1}
shows the corresponding ratios 
$$
\frac{{\rm direct}}{{\rm direct+fragmentation}}\; ,
\;\;\;\;\;\frac{{\rm fragmentation}}{{\rm direct+fragmentation}} \; ,
$$
as functions of the photon transverse momentum $p_T$, for
$\sqrt{S}=31.5$~GeV, corresponding to a typical fixed-target
energy.
Here we have considered $pp$ collisions, and
we have chosen the factorization/renormalization
scales as $\mu=p_T$. One can see that the NLO fragmentation component
contributes about $40\%$ of the cross section at the lowest
$p_T$ shown and then rapidly decreases, becoming lower than
$10\%$ at $p_T\approx 11$~GeV. As we anticipated earlier, 
threshold resummation affects the fragmentation component
much more strongly than the direct part. After resummation,
the fragmentation contribution is relatively much more important,
as shown in Fig.~\ref{fig1}, yielding almost $60\%$ of the cross
section at smaller $p_T$ and still more than $20\%$ at 
$p_T= 11$~GeV. 
 
\begin{figure}[t!]
\begin{center}
\epsfig{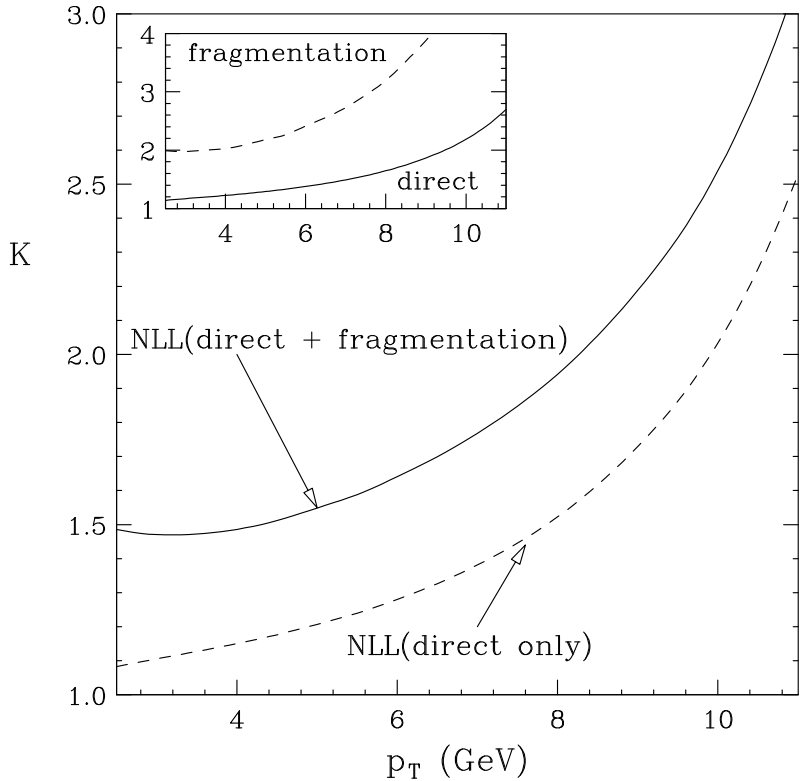}
\end{center}
\vspace*{-.5cm}
\caption{``$K$-factors'' as defined in Eq.~(\ref{kfac})
for the case where only the direct component is resummed,
and for the case where NLL resummation is applied to both the direct and 
the fragmentation contributions. Parameters are as for Fig.~\ref{fig1}.
The insert shows the individual ``$K$-factors'' for the direct
and the fragmentation pieces.
\label{fig2} }
\vspace*{0.cm}
\end{figure}

Similar conclusions are reached when one analyzes the additional 
enhancement that NLL resummation gives over NLO. In Fig.~\ref{fig2}
we show the ``$K$-factors''
\begin{equation}
K\equiv \frac{d\sigma^{\rm (match)}}{d\sigma^{\rm (NLO)}} \label{kfac}
\end{equation}
for the case where only the direct contribution is resummed
(and the fragmentation one taken into account at NLO), and 
for the case when both contributions, direct and fragmentation,
are resummed. We have chosen the same energy and other
parameters as in the previous figure. In agreement with earlier
studies~\cite{CMNOV,KO,sv}, 
resummation of the direct contribution alone is fairly
unimportant at lower $p_T$, yielding a ``$K$-factor'' close 
to unity. In contrast to this, taking into account the NLL 
resummation of the fragmentation component as well leads 
to a much bigger ``$K$-factor'', roughly a $50\%$ enhancement 
over NLO at the lower $p_T$, and even a factor 2.5$-$3 at the
highest $p_T$ considered. The insert in the figure shows
the individual  ``$K$-factors'' for the direct and the 
fragmentation components. The one for the fragmentation 
piece is very large, albeit not as large as what was found
for the case of $\pi^0$ production in our previous study~\cite{ddfwv}.
This finding is explained by the fact that gluonic channels
receive much larger resummation effects than quark ones, but 
that the such channels are relatively suppressed in the photon production 
case since the gluon-to-photon fragmentation function is 
much smaller than the gluon-to-pion one. 

\begin{figure}[t!]
\begin{center}
\epsfig{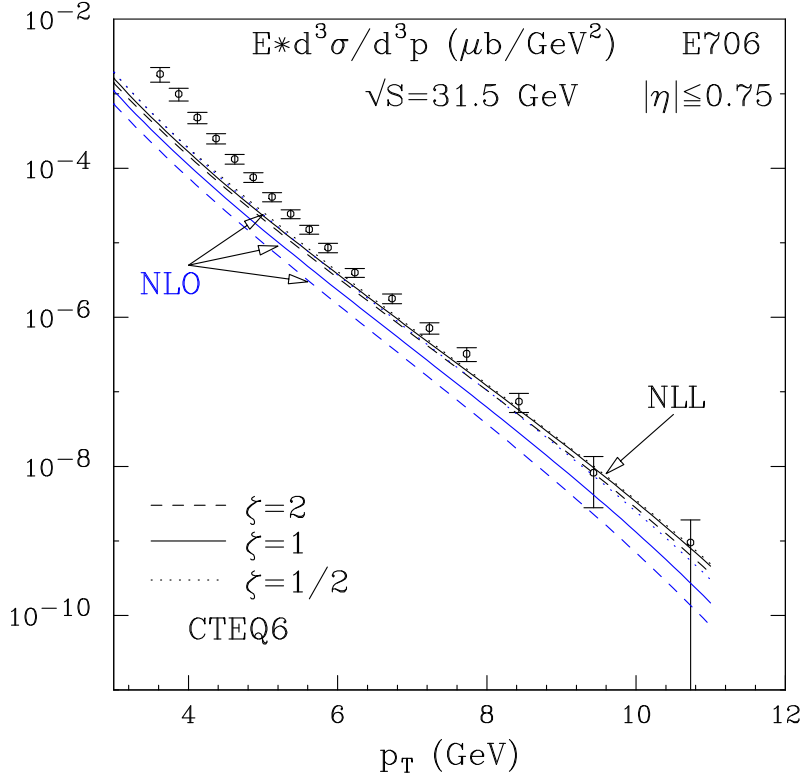}
\end{center}
\vspace*{-.5cm}
\caption{Comparison of NLO and NLL resummed calculations of the cross
section for $pBe\to\gamma X$ to data from E706~\cite{e706},  
at $\sqrt{S}=31.5$~GeV and $|\eta|\leq 0.75$, for three different 
choices of the renormalization/factorization scale $\mu$.
\label{fig3} }
\vspace*{0.cm}
\end{figure}

\begin{figure}[t!]
\begin{center}
\epsfig{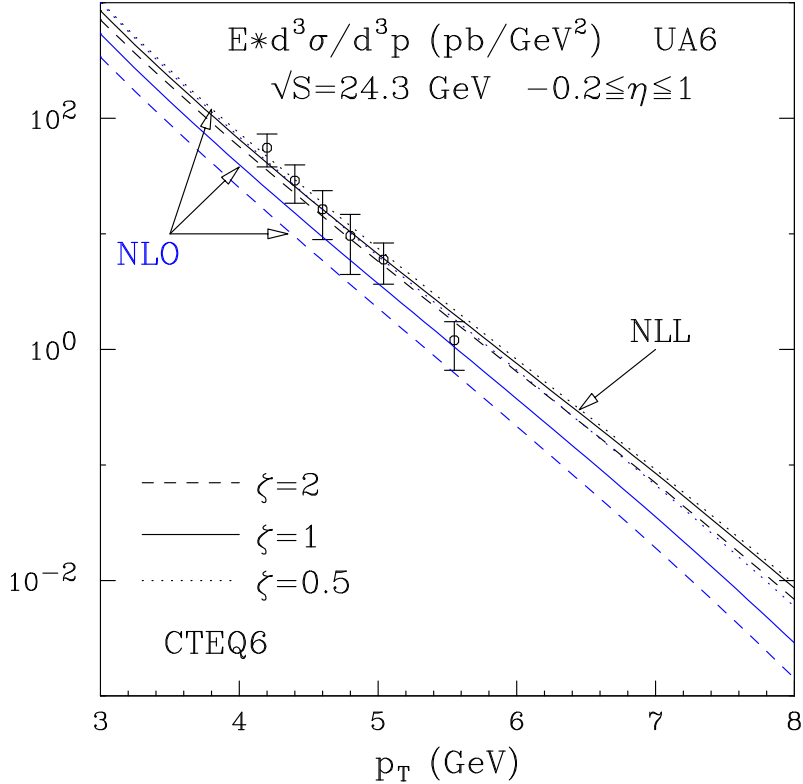}
\end{center}
\vspace*{-.5cm}
\caption{Same as Fig.~\ref{fig3}, but comparing to data from
UA6~\cite{ua6} for $pp\to\gamma X$ at $\sqrt{S}=24.3$~GeV 
with $-0.2\leq\eta\leq 1$. 
\label{fig4} }
\vspace*{0.cm}
\end{figure}

\begin{figure}[t!]
\begin{center}
\epsfig{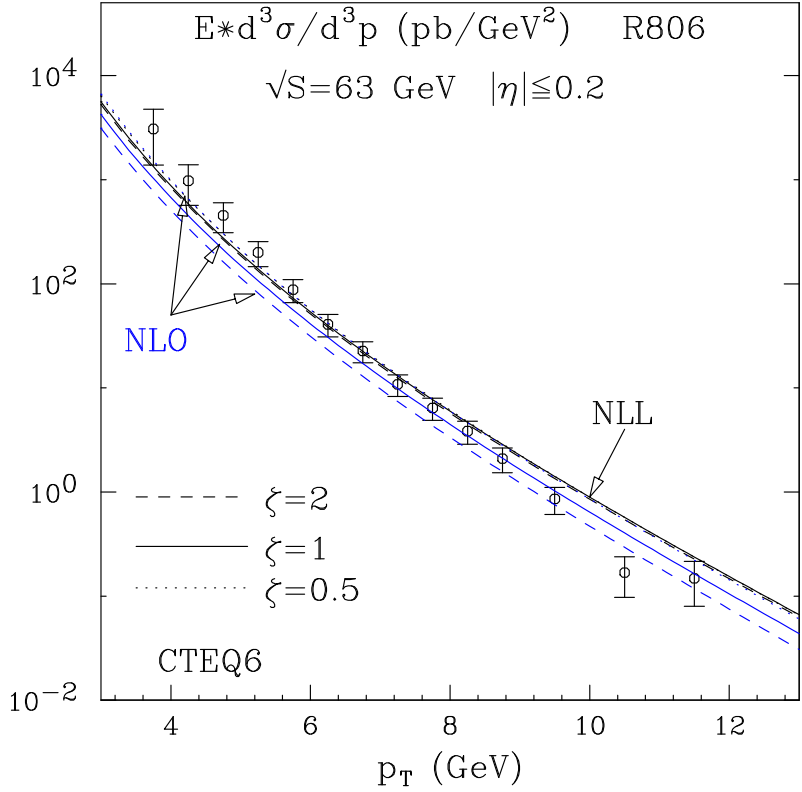}
\end{center}
\vspace*{-.5cm}
\caption{Same as Fig.~\ref{fig3}, but comparing to data from
R806~\cite{r806} for $pp\to\gamma X$ at $\sqrt{S}=63$~GeV with 
$|\eta|\leq 0.2$.  
\label{fig5} }
\vspace*{0.cm}
\end{figure}

From Fig.~\ref{fig2} we may conclude that NLL resummation
of the fragmentation component leads to a significant enhancement
of the theoretical prediction and will have some relevance
for comparisons of data and theory. Such comparisons
are shown in Figs.~\ref{fig3}-\ref{fig5}. In  Fig.~\ref{fig3}
we show the data for $pBe\to\gamma X$ from the E706 experiment~\cite{e706},
along with our theoretical calculations at NLO and for the 
NLL resummed case. The energy is $\sqrt{S}=31.5$~GeV, as used
for the previous figures, and the data cover $|\eta|\leq 0.75$. We
give results for three different choices of scales, $\mu=
\zeta p_T$, where $\zeta=1/2, 1, 2$. It is first of all evident from the
figure that the NLO result falls far short of the data. As we 
shall see below, this shortfall is particularly pronounced for
the E706 data. Furthermore, there is a very large scale dependence 
at NLO. When the NLL resummation is taken into account, the scale 
dependence is drastically reduced. This observation was already
made in the previous phenomenological studies of the resummed 
prompt-photon cross section~\cite{CMNOV,KO,sv}, 
in which however only resummation for 
the direct component was implemented. As can also be seen from 
Fig.~\ref{fig3}, at the lower $p_T$ the full resummed result is 
roughly at the upper end of the ``band'' generated by the scale 
uncertainty at NLO, whereas at the higher $p_T$ it is considerably
higher. Overall, as we saw in Fig.~\ref{fig2}, there is a further 
significant enhancement over previous NLL resummed results~\cite{CMNOV,KO,sv}. 
This additional enhancement leads to a 
moderate improvement of the comparison between theory and the
E706 data. Clearly, even with NLL resummation of the 
fragmentation component the calculated cross section remains
far below the E706 data, except for $p_T\gtrsim 8$~GeV.

Figure~\ref{fig4} shows similar comparisons with the 
data for $pp\to \gamma X$ from UA6~\cite{ua6} at $\sqrt{S}=24.3$~GeV. 
Here, the resummed
calculation, which again shows a very small scale dependence, is in very
good agreement with the data. As before, resummation of the fragmentation
component leads to a non-negligible enhancement of the cross section,
pushing the theoretical NLL results to or slightly beyond the upper end
of the NLO scale uncertainty band. Finally, in Fig.~\ref{fig5} we 
show R806 results for $pp\to \gamma X$ from the ISR at $\sqrt{S}=63$~GeV. 
Similar features as before are observed. Note that we are further away from
threshold here, due to the higher energy of the ISR. It is likely that 
the NLL resummation is not completely accurate here, but that 
terms subleading in $N$ could have some relevance. We reserve the 
closer investigation of this issue to a future study. 

\noindent
{\bf Conclusions and outlook.} We have studied the NLL all-order 
resummation of threshold logarithms in the partonic cross sections 
relevant for high-$p_T$ prompt-photon production. The novel 
feature of our study is that we have also taken into account
the NLL resummation of the photon fragmentation component. 
Here we were motivated by the rather large enhancements
that we had found in a previous study of threshold resummation
for the process pp$\to \pi^0X$. The theoretical description for this 
process is the same as that for the fragmentation component 
to the prompt photon cross section; the only difference arises
in the use of pion vs. photon fragmentation functions. 

We have found that indeed the fragmentation component 
is subject to much larger resummation effects than the
direct one. This implies that probably a substantially 
larger fraction of observed  
photons than previously estimated are produced in jet fragmentation.
On the other hand, we also found that the enhancement
of the fragmentation component due to the threshold logarithms
is smaller than the enhancement previously observed for
$\pi^0$ production, mostly as a result of the smallness of the 
photon-to-gluon fragmentation function as compared to the
gluon-to-pion one. We note, however, that fairly little is
known about the function $D^{g\to \gamma}$. It is probably not 
ruled out that this functions is much bigger than expected
in the set~\cite{grvf} of photon fragmentation functions
that we have used, in which case resummation effects would
become yet more substantial. 
 
The fully resummed prompt-photon cross section shows
a much reduced scale dependence. We find that the comparison
of the NLL resummed cross section with experimental data
shows varied success, with the theoretical calculations still
lying much lower than the E706 data, but consistent
with the UA6 and R806 $pp$ data. In the light of this, further studies 
and more detailed comparisons are desirable. We note that generally 
any residual shortfall of the resummed theoretical results would 
likely need to be attributed to non-perturbative contributions that are
suppressed by inverse powers of the photon transverse
momentum. These could for example be related to small ``intrinsic''
parton transverse momenta~\cite{ktgamma}. Resummed perturbation 
theory itself may provide information on the structure 
of power corrections, through contributions to the resummed 
expressions in which the running coupling constant is probed at 
very small momentum scales. A recent study~\cite{gswv} addressed 
this issue in the case of the prompt-photon cross section at large 
$x_T$ and estimated power corrections to be not very sizable. 

Our study improves the theoretical description and thus is a step 
towards a better understanding of the 
prompt-photon cross section in the fixed-target regime.

\noindent
{\bf Acknowledgments.} The work of D.dF has been partially 
supported by Conicet,
Fundaci\'on Antorchas, UBACyT and ANPCyT.
W.V.\ is grateful to RIKEN, Brookhaven National Laboratory
and the U.S.\ Department of Energy (contract number DE-AC02-98CH10886) for
providing the facilities essential for the completion of his work.

\end{document}